\documentclass{article}
\usepackage{makeidx,epsfig}
\usepackage{setspace,graphicx}%,subfigure
\usepackage{multirow} % here
\usepackage{amsfonts}
\usepackage{pst-all}
\newtheorem{theorem}{Theorem}
\newtheorem{definition}[theorem]{Definition}
\newtheorem{lemma}[theorem]{Lemma}
\def\cocoa{{\hbox{\rm C\kern-.13em o\kern-.07em C\kern-.13em o\kern-.15em A}}}
\newtheorem{example}{Example}

\title{The algebraic method in experimental design}
\author{Hugo Maruri-Aguilar and Henry P. Wynn}

\begin{document}
\maketitle
%\tableofcontents

\abstract{The algebraic method provides useful techniques to identify models
in designs and to understand aliasing of polynomial models. The present note
surveys the topic of Gr\"obner bases in experimental design and then describes
the notion of confounding and the algebraic fan of a design. The ideas are
illustrated with a variety of design examples
ranging from Latin squares to screening designs.}

\tableofcontents
%\chapter
%{The algebraic method in experimental design}

\section{Ideals and varieties: introducing algebra}
We are familiar with the use of polynomials throughout statistics.
For example, much of this handbook is concerned with design for
polynomial regression. Thus we have polynomial terms in factors
$x_1,x_2,\ldots$:
$$x_1, x_2^2, x_1x_2,\ldots$$
and a second order polynomial response surface  in two factors is
\begin{equation}\label{ch11_ex1}f(x_1,x_2) = \theta_{00} + \theta_{10} x_1 + \theta_{01} x_2 + \theta_{20}x_1^2+ \theta_{11}x_1x_2+\theta_{02}x_2^2.\end{equation}

The first, but very important,  algebraic point is that polynomials
are made up of linear combinations of monomials. Consider a set of $k$ factors
$x_1, \ldots, x_k$ and non-negative integers $\alpha = (\alpha_1,\ldots,\alpha_k)$;
a monomial is
$$x^{\alpha}= x_1^{\alpha_1}x_2^{\alpha_2} \cdots x_k^{\alpha_k}.$$
Note that when we use the term polynomial we shall typically mean a
polynomial in one {\em or more} variables.

A monomial  $x^{\alpha}$ can be represented by its exponent vector
$\alpha$ and we can list the monomials in a model either directly or
by listing a set of exponents. We shall often use the notation
$\{x^{\alpha}, \alpha \in M\}$, for some set of exponents, $M$. This
chapter is largely concerned with the interaction between the choice
of a design and the list $M$. We know from classical factorial
design that only some models are estimable for a given design and so
any such theory must be intimately related to the problem of aliasing
and we shall cover this is section \ref{ch11_alias}.

The set of all polynomials over a base field is a ring, so that
rings are the basis of the theory. Thus, given a base field $K$ we
obtain the ring of polynomials, $R=K[x_1, \ldots, x_k]$ over $K$,
which are linear combinations of monomials with coefficients in the
base field. Our ``$\alpha$" notation allows us to write this
compactly as
$$f(x) = \sum_{\alpha \in M} \theta_{\alpha} x^{\alpha},$$
where, as above, $M$ is a finite set of distinct exponents and
clearly $f(x)\in R$. For
example, the set $M$ for the polynomial model in
Equation (\ref{ch11_ex1})
is $\{(0,0),(1,0),(0,1),(2,0),(1,1),(0,2)\}$.

Given that we have launched into algebra we need to introduce the
first two essentials: {\em ideals} and {\em varieties}. In what
follows we present only the basic ideas of the theory, pointing the
reader to \cite{CLO1996} or \cite{PRW2000} for further details.

For a ring $R$ we have special subsets called {\em ideals}.
\begin{definition}
A subset $I \subset R$ is an ideal if for any $f,g \in I$ we have
$f+g \in I$ and for any $f \in I$ and $g \in R$ we have $fg \in I$.
\end{definition}
The  ideal generated by a finite set of polynomials
$\{f_1,\ldots,f_m\}$ is the set of all polynomial
combinations:
$$\langle  f_1,\ldots,f_m\rangle = \{f_1g_1+ \cdots + f_mg_m:\; g_1,\ldots,g_m \in R\}$$

To have some immediate intuition consider a single point $x$. The
set of all polynomials $f$ such $f(x) = 0$ is an ideal: since for
any polynomial  $g(x)$ if $f(x)=0$ we have $g(x)f(x)=0$. In the next section
this will be extended to sets of points, namely designs. The Hilbert
basis theorem says that any (polynomial) ideal $I$ is finitely
generated, i.e. for any ideal we can find a finite collection
$f_1,\ldots,f_m \in R$ such that $I\;=\;\langle
f_1,\ldots,f_m\rangle$.

We are familiar with linear varieties expressed by setting some
linear polynomial function equal to zero. Thus a straight line can be
written as the collection of points $(x_1,x_2)$ in two dimensions
such that $ax_1 + bx_2 + c = 0,$
for constants $a,b,c$. An algebraic variety is the extension of this
concept to simultaneous solutions of a set of polynomial equations.

\begin{definition}
Let $f_1, \ldots, f_m \in K[x_1,\ldots, x_k]$ be a set of
polynomials. The associated affine variety is the solution (also
called {\em zero set}) of a set of simultaneous equations they
define:
$$V(f_1, \ldots, f_n) = \{(a_1,\dots,a_k) \in K^k: f_i(a_1, \ldots,
a_d)=0,\; i=1, \dots, m\}$$
\end{definition}
Every affine variety has an associated ideal which we write $I(V)$.
It is the set of {\em all} polynomial which are zero on the variety:
$$I(V) = \{f \in K[x_1,\ldots, x_k] :
f(a_1,\ldots,a_k)=0,\;\mbox{for all}\; (a_1,\ldots,a_k) \in V\}.$$

What appears to be a straightforward relationship between ideals and
varieties is actually very subtle. If we start with polynomials
$f_1, \ldots, f_m$ and  construct the corresponding variety $V$ and
form the ideal $I(V)$, is it true that $I(V) = \langle  f_1, \ldots,
f_m \rangle$? We can always claim that $\langle  f_1, \ldots, f_m
\rangle\;\subset \; I(V)$, but the converse may not be true and
refer to \cite{CLO1996} for a detailed discussion. Fortunately, for
a design, the variety is collection of isolated single points,
the equivalence holds and we may move freely between ideals and
designs.

\section{Gr\"obner bases}\label{sec_gb}

Perhaps the most important construction in abstract algebra is that
of a {\em quotient}. Give two polynomials $f,g \in
K[x_1,\ldots,x_k]$ and an ideal $I$ define the equivalence class $f
\sim_I g$ if and only if $f-g \in I$. The members of the quotient
$K[x_1, \ldots x_n]/I$ are the equivalence classes. Since $f_1
\sim_I f_2$ and $g_1 \sim_I g_2$ imply $f_1+g_1 \sim_I f_1+g_2$ and
$f_1g_1 \sim_I f_2g_2$, then $K[x_1, \ldots x_n]/I$ is also a ring.
Finding $K[x_1, \ldots x_n]/I$ in a particular case requires a {\em
division algorithm}. Finding  a quotient computationally needs a
division algorithm.

\subsection{Term orderings}

Let us recall division of polynomials in one dimension. If we divide
$1 + 3x + 2x^2 + x^3$ by $2+x$ we would obtain the tableau:
\label{ch11_divex}
$$\begin{array}{rll}
      &\;\;\;\; x^2+3    \\
  x+2&\multicolumn{2}{c}{ \overline{)\; x^3+2x^2+3x+1} }\\
   & \;\;-x^3-2x^2   \\\cline{2-3}
   && \;\;3x+1   \\
   &&  -3x-6  \\\cline{3-3}
   &&\multicolumn{1}{r}{-5} \\
\end{array}$$
giving: $x^3+2x^2+3x+1 = (x^2+3)(x+2)-5$. We give this example to
remind ourselves that at each stage we need to use the leading term. To
obtain leading term we need an ordering. In one dimension the
ordering is
$ 1 \prec x \prec x^2 \prec \cdots,$
That is, we order by degree and division is unique.  This is
generalised to a special total ordering on monomials
$\{x^{\alpha}\}$.
\begin{definition}
A {\em monomial term ordering}, $\prec$, is a total ordering of
monomials such that $1 \prec x^{\alpha}$ for all $\alpha \geq 0$,
$\alpha \neq 0$ and, for all $\gamma \geq 0$, $x^{\alpha} \prec x^{\beta}$ implies
$x^{\alpha +\gamma} \prec x^{\beta +\gamma}$.
\end{definition}

We shall use the term {\em monomial ordering} for short.
There is a number of standard monomials orderings.
\begin{enumerate}
\item {\em Lexicographic ordering, Lex}.
$x^\alpha \prec_{Lex} x^\beta$ when (i) $\beta - \alpha \geq 0$ and the
leftmost entry of $\beta - \alpha$ is positive.
\item {\em Graded lexicographic ordering, DegLex}: $x^\alpha \prec_{DegLex} x^\beta$
if (i) the degree of $\alpha$ is less than that of $\beta$,
$|\alpha|  < |\beta|$ and (ii) $\alpha \prec_{Lex} \beta$
\item {\em Reverse lexicographic ordering, DegRevLex}: $x^\alpha \prec_{DegrevLex} x^\beta$
if (i) $|\alpha|
< \beta|$ and (ii) $\overline{\alpha} \prec_{Lex} \overline{\beta}$,
where the overline means: reverse the entries.
\end{enumerate}

 Graded orderings are orderings in which the
 first comparison between monomials is determined by their total degree.
  For example, under a graded order, $x_i^2\succ x_j$ for any indeterminates
 $x_i,x_j$ in the ring $K[x_1,\ldots,x_k]$. The degree lexicographic and
 degree reverse lexicographic term orders above fall in this class.
Contrary to graded orderings, for a lexical ordering in which $x_i\succ x_j$
then $x_i\succ x_j^m$ for $m=1,2,\ldots$ thus making all powers of $x_j$ lower
than $x_i$.

\subsection{Matrix based term orderings}

Monomial term orderings can be defined using products with
 matrices and element-wise comparisons.  If
the exponents of monomials $x^\alpha,x^\beta$ are considered
as row vectors, we say that
$x^\alpha\prec_M x^\beta$ if $M\alpha^T<M\beta^T$, where
$M$ is a non-singular matrix and
the inequality is tested element-wise starting from the first element.
The matrix $M$ above
satisfies certain conditions which are stated in the following
theorem \cite{PRW2000}.

\begin{theorem}Let $M$ be a full rank matrix of size $k\times k$ such
that the first non-zero entry in each column is positive. Then
$M$ defines a term ordering in the following sense:
\begin{enumerate}
\item For every vector $\alpha\in\mathbb{Z}^k_{\geq 0}$ with
 $\alpha\neq (0,\ldots,0)$ then $(0,\ldots,0)<M\alpha^T$ and
\item For any pair of vectors $\alpha,\beta,\gamma
\in\mathbb{Z}^k_{\geq 0}$
such that $M\alpha^T<M\beta^T$ then
$M(\alpha+\gamma)^T<M(\beta+\gamma)^T$.
\end{enumerate}
%where the inequalities are performed element-wise
%starting from the first element of vectors.
\end{theorem}

The identity matrix of size $k$ corresponds to the lexical
term ordering. Note that the relation between ordering matrices
and term orderings is not a one to one. A matrix $M'$ defining the
same ordering as $M$ can be obtained by multiplying each row
of $M$ by a positive constant so for instance the matrix with
diagonal $1,2,\ldots,k$ and zeroes elsewhere also defines a lexical
term ordering.
Usually only integer entries are used for computations although
the theory does not preclude using for instance, matrices with
rational or real entries \cite{CLO1996}.

 An important case of ordering matrices is that of matrices for
 graded orderings. Any full rank matrix $M$ in which all elements
 of the first row are a positive constant defines a graded ordering.
The degree lexicographic term ordering is built with
a matrix $M$ with all entries one in its first row
and the remaining rows are the top $k-1$ rows of an identity matrix.
The \cocoa{ } command \texttt{Use T::=Q[x,y,z], DegLex;} creates the
same ring and ordering when the matrix and ring are defined with the commands

\texttt{M:=Mat([[1,1,1],[1,0,0],[0,1,0]]);}

\texttt{Use T::=Q[x,y,z], Ord(M);}

The querie \texttt{xy\^{}2>x\^{}2z;}
  yields output \texttt{FALSE} which means that $xy^2\prec x^2z$
  under the graded lexicographic order in which $x\succ y\succ z$.

The standard ordering in the software system \cocoa{ } is the
degree reverse lexicographic (\texttt{DegRevLex}), which is
implicit in the following ring definition

\texttt{Use T::=Q[x,y,z];}

\texttt{xy\^{}2>x\^{}2z;}

The output of
the querie is \texttt{TRUE} and this is interpreted as
$xy^2\succ x^2z$ under a degree reverse lexicographic term
ordering in which $x\succ y\succ z$. Note the reversal of the
ordering between the two monomials for the previous graded order.

A more specialized and efficient instance of matrix
orderings is produced by a using a single row matrix,
in which case we say ``ordering vector''.
An ordering vector defines only a partial but not a total ordering
over $R$. For example the vector $w=(1,1,1)$ naturally produces
the ordering $xy^2\succ_w xz$ because \[(1,1,1)(1,2,0)^T=3>2=(1,1,1)(1,0,1)^T,\]
yet it cannot distinguish between monomials of the same degree
such as $xy^2$ and $z^3$.
However, Gr\"obner basis
are computed over finite sets of monomials
rather than over all monomials with exponents in
$\mathbb{Z}^k_{\geq 0}$. This last fact together with
a careful selection
of the ordering vector are at the core of the efficient Universal
Gr\"obner bases algorithms \cite{BOT2003,MA2006}.

\subsection{Monomial ideals and Hilbert series}

Now that we have a total ordering any finite set of monomials has a leading
term. In particular, since a polynomial, $f$, is based on a finite
set of monomials it has a unique leading term. We write it
$LT_{\prec} (f)$, or, if $\prec$ is assumed, just $LT(f)$.

A monomial ideal is an ideal generated by monomials. Monomial ideals
play a critical part in computational methods for polynomials.
\begin{definition}
A monomial ideal $I$ is an ideal for which  a collection of
monomials $f_1,\ldots, f_m$ such that any $g \in I$ can be expressed
as a sum $$g = \sum_{i=1}^m g_i(x)f_i(x).$$
\end{definition}
Multiplication of monomials is just achieved by adding exponents:
$$x^{\alpha} x^{\beta} = x^{\alpha + \beta},$$
and $\alpha + \beta$ is in the positive (shorthand for non-negative)
``orthant" with corner at $\alpha$. The set of all monomials in a
monomial ideal is the union of all positive orthants whose corners
are given  by the exponent vectors of  the generating monomial $f_1,
\ldots, f_m$.

For a given monomial ideal, a complete degree by degree description of
the monomials inside the ideal or, equivalently, those outside the ideal
is given by the Hilbert function and series. Here we only give the basic
idea, referring the reader to references \cite{CLO1996} and \cite{CLO1998}
for a full description.

\begin{definition}
Let $I$ be a monomial ideal in $R$.
\begin{enumerate}\item For all non-negative degrees $j$,
the Hilbert function $HF_I(j)$ is the number of
monomials not in $I$ of total degree $s$.
\item the Hilbert series of $I$ is the formal series
$HS_I(s)=\sum_{j=0}^\infty s^jHF(j)$.\end{enumerate}
\end{definition}
The Hilbert series is the generating function of the
Hilbert function. Both count monomials which are not in
the monomial ideal $I$. In what follows, unless it is required, we omit the
subindex referring to the monomial ideal.

\begin{example}\label{ex_hs}{\normalfont Consider the ideal
$I=\langle x^3,xy^2,y^4\rangle\subset k[x,y]$. The monomials
which do not belong to $I$ are
$1, x, x^2, y, yx, yx^2, y^2, y^3$, so the Hilbert function
equals $1,2,3,2$ for $j=0,1,2,3$ and zero for all $j\geq 4$.
The Hilbert series is thus $HS(s)=1+2s+3s^2+2s^3$. }\end{example}

Note that monomials in the first orthant are
counted with the formal series $\sum_{j=0}^\infty {j+k-1\choose j}s^j$
where $k$ is the number of indeterminates. Then, by substracting
the Hilbert series $HS(s)$ from the last expression we have a generating
function to count monomials inside $I$.
Using two dimensions as in
Example \ref{ex_hs}, the formal series for the firtst orthant is 
\[\sum_{j=0}^\infty {j+1\choose j}s^j=
\sum_{j=0}^\infty(j+1)s^j=\frac{1}{(1-s)^2},\] which counts
the monomials in the first quadrant.
The generating function for the number of monomials in $I$
for each degree is found by substraction:
\[\frac{1}{(1-s)^2}-(1+2s+3s^2+2s^3)=\frac{2s^3+s^4-2s^5}{(1-s)^2}.\]
The alternating signs in the polynomial in the numerator are related
to inclusion-exclusion rules and although it may seem a simple
calculation above, in general
determining a simple form of the numerator in this last computation
 is not a simple task.

\begin{example}{\normalfont
Consider the monomial ideal in $k[x_1,\ldots,x_7]$ generated by
monomials $x_1^2,\ldots,x_7^2$ and by all pairs $x_ix_j$, $1\leq i<j\leq 7$.
This ideal has Hilbert Function with values $1$ and $7$ for $j=0,1$ and zero for
$j\geq 2$ so its Hilbert Series is $HS(s)=1+7s$, i.e. one monomial of degree
zero and seven
monomials of degree one outside the ideal.
The monomials outside this ideal are $1,x_1, x_2, x_3, x_4, x_5, x_6, x_7$,
which later will be understood as a model for the
Plackett-Burman design in Examples \ref{ex_pb}
and \ref{ex_pb2}. See also first row of
Table \ref{tab_pb8}.
%$I=\langle
%x_1^2, x_2^2, x_3^2, x_4^2,  x_5^2,  x_6^2, x_7^2
%x_1x_2, x_1x_3, x_1x_4, x_1x_5, x_1x_6, x_1x_7,
%x_2x_3, x_2x_4, x_2x_5, x_2x_6, x_2x_7,
%x_3x_4, x_3x_5, x_3x_6, x_3x_7,
%x_4x_5, x_4x_6, x_4x_7,
%x_5x_6, x_5x_7,
%x_6x_7, \rangle$
}\end{example}

\subsection{Gr\"obner bases}

Dickson's Lemma states that, even if we define a monomial ideal with
an infinite set of $f_i$, we can find a finite set $h_1, \ldots h_n$
such that $I = \langle  h_1,\ldots, h_k \rangle$.  But there are, in
general, many ways to express a ideal $I$ as being generated from a
basis $I = \langle f_1,\dots, f_m\rangle$.

\begin{definition}
Given an ideal $I$ a set $\{g_1, \ldots g_m\}$ is called a Gr\"obner
basis if:
$$\langle LT(g_1), \ldots, LT(g_m)\rangle\; = \; \langle LT(I)\rangle,$$
where $\langle LT(I)\rangle$ is the ideal generated by all the
monomials in $I$.
\end{definition}
We sometimes refer to $\langle LT(I) \rangle$ as the {\em leading
term ideal}.
\begin{lemma}
Any ideal $I$ has a  Gr\"obner basis and any Gr\"obner basis in the
ideal is a basis of the ideal.
\end{lemma}

Monomial orderings are critical in establishing that for any given
monomial ordering,  $\prec$, any ideal $I$ has a unique ``reduced"
Gr\"obner basis. Given a monomial ordering, $\prec$, and an ideal
expressed in terms of the G-basis, $I\;=\; \langle g_1,\ldots,
g_m\rangle$ with respect to that monomial ordering any polynomial
$f$ has a unique remainder, $r(x)$ with respect the quotient
operation $K[x_1, \ldots, x_k]/I$. That is
\begin{equation}\label{polydiv}f = \sum_{i=1}^m s_i(x)g_i(x) + r(x)\end{equation}
We call the remainder $r(x)$ the {\em normal form} of $f$ with
respect to $I$ and write $r(x) =NF(f)$. Or, to stress the fact that
it may depend on $\prec$, we write $NF(f, \prec)$.

The division of a polynomial in Equation (\ref{polydiv}) is the
generalization of simple polynomial division such as that of the
example shown in Page \pageref{ch11_divex}, where the result
was $s_1(x)=x^2+3$ with remainder $r=-5$. In other words
the normal form of $1 + 3x + 2x^2 + x^3$ with respect to
the ideal generated by $g_1(x)=2+x$ is $-5$.

Here are some formal
definitions.
\begin{definition}
Given a monomial ordering $\prec$, a polynomial $f=\sum_{\alpha \in
L} \theta_{\alpha} x^{\alpha}$ is a normal form with respect to
$\prec$ if $x^{\alpha} \notin \langle LT(f) \rangle $ for all
$\alpha \in L$.
\end{definition}

\begin{lemma}
Given an ideal $I$  and a monomial ordering $\prec$, for every $f
\in K[x_1,\ldots,x_k]$ there is a unique normal form $NF(f)$ such
that $f-NF(f) \in I$.
\end{lemma}

We now need to relate (i) the Gr\"obner basis, (ii) a division
algorithm and (iii) the nature of the normal form. We have partly
covered this but let us collect the results together.
\begin{enumerate}
\item There are algorithms, which  given an ideal, $I$, a
monomial ordering $\prec$ and a polynomial $f$ deliver the remainder
$r$, in Lemma (8), by successively dividing by the G-basis terms
$g_i,\;i=1,\dots,m$. The best known is the Buchberger algorithm.
\item Suppose the remainder $r(x)=NF(f) = \sum_{\alpha \in L}\theta_{\alpha} x^{\alpha}$,
then $\{x^{\alpha}, \alpha \in L\}$ is precisely the set of
monomials {\em not} divisible by any of the leading terms of the
G-basis of $I:\; \mbox{LT}(g_i),\; i=1, \ldots, m$.
\item The remainder $r(x)= NF(f)$ does not depend on which order the
G-basis terms $g_i(x)$ are used in the division algorithm.
\item The (maximal) set $\{x^{\alpha} , \alpha \in L\},$ which can appear in a  remainder
$r(x)$ is a basis of the quotient ring, considered as a vector space
of functions over $k[x_1, \ldots, x_k]/I$. The terms are linearly
independent over $I$:
$$\sum_{\alpha}\theta_{\alpha}x^{\alpha} \sim_I 0$$
implies $\theta_{\alpha} =0$ for all $\alpha \in L$.
\end{enumerate}

\subsection{Software tools}
All the operations defined above are available on modern computer
algebra software. Here is a brief list, the full list is very
extensive and extends to nearly all areas of computer algebra,
sometimes called computational algebraic geometry:
\cite{cocoa}\cocoa, \cite{macaulay}, \textit{macaulay2}, \cite{gfan}
\texttt{gfan}, \cite{singular} \texttt{Singular}. A rough list of
capabilities relevant to this chapter is as follows.
\begin{enumerate}
\item Construction of monomial orderings; the standard ones are
usually named and immediately available
\item Ideal operations such as unions, intersections,
elimination
\item Buchberger algorithm and modern improvements, quotienting,
Normal Forms
\item Special algorithms for {\em ideals of point}. We shall use
these extensively in our examples
\item Gr\"obner fan. See section \ref{sec_fans}
\end{enumerate}

\section{Experimental design}\label{sec_ed}
We have indicated already that for applications to design we should think
of design as lists of points,
$$D = \{x^{(1)}, \ldots x^{(n)}\},$$
in $R^k$. As algebraic varieties they have associated ideal
$$I(D) = \{f : f(x) = 0, \; x \in D\}$$

The use of polynomials to define design is clearly not new. For
example a  $2^k$ full factorial  designs give by $\{\pm 1, \ldots,
\pm 1\}$ is expressed the solution of the simultaneous equations:
$$\{ x_i^2 -1 = 0,\; i = 1, \ldots, k\}.$$
To obtain fractions we impose additional equations: e.g. $x_1\ldots
x_k = 1$.

We now give what can loosely be described as the {\em algebraic
method} in the title of this Chapter. We do this in a step-by-step
approach.

\begin{enumerate}
\item Choose a design $D$
\item Select a monomial term ordering, $\prec$
\item Compute Gr\"{o}bner basis for $I(D)$ for  given monomial
ordering, $\prec$.
\item The quotient ring $$K[x_1,\ldots,x_k]/I(D)$$ of the ring of polynomials
$K[x_1,\ldots,x_k]$ in $x_1,\ldots, x_k$ forms is a vector space
spanned by a special set of monomials: $x^{\alpha}, \alpha \in L$.
These are all the monomials not divisible by the leading terms of
the G-basis and $|L|= |D|$.
\item The set of multi-indices $L$ has the ``order ideal"
property: $ \alpha \in L$ implies $\beta \in L$ for any $0 \leq
\beta  \leq \alpha$. For example, if $x_1^2x_2$ in the model so is
$1,x_1,x_2,x_1x_2$.
\item Any function $y(x)$ on $D$ has a unique  polynomial
interpolator given by
$$ f(x) = \sum_{\alpha \in L} \theta_{\alpha} x^{\alpha} $$
such that $y(x)= f(x),\; x \in D$.
\item The cardinality of the design and the quotient basis is the same: $|L|= |D|$.
\item The $X$-matrix is $n \times n$, has full rank $n$ and  has rows
indexed by the design points and columns indexed by the basis:
$$X = \{x^{\alpha}\}_{x \in D, \alpha \in L}$$
\end{enumerate}

The implications of the method are considerable. But at its most
basic it says that {\em we can always find a saturated polynomial
$f(x)$ interpolating data over an arbitrary design} $D$.

The shape of the model index set $L$ arising from the order ideal
property is important. It is exactly the shape which, in the
literature has been called variously: ``staircase models'',
``hierarchical models'', ``well-formulated models'', or
``marginality condition'', see \cite{N1977,P1990}.  It can be be
seen easily from the fact that the multi-index terms given by $L$
are the complement in the non-negative integer orthant of those
given by the monomials in the monomial ideal of leading terms: the
complement of a union of orthants has the staircase property.

We now give a number of examples.

\begin{example}{\normalfont \label{ex_bj} Screening designs.
A class of designs for main effect estimation while simultaneously avoiding
biases caused by the presence of second order effects and avoid confounding
of any pair of second order effects was recently proposed \cite{JN2011}. The
authors produced designs of size $n=2k+1$ for different dimensions ranging
from $k=4$ up to $k=30$,
and their construction is based on folding a certain small fraction of size $k$
of a $3^k$ design with levels $-1,0,1$ and then adding the origin.
Naturally that after folding and adding the origin, the
screening design still remains
a special fraction of $3^k$ design.
Here we consider the designs for $k=7$ and $k=10$ in Table \ref{tab_bjd}.

% Table 1
\begin{table}[h]\begin{small}
$\begin{array}{rrrrrrr}
x_1&x_2&x_3&x_4&x_5&x_6&x_7\\\hline
0&1&-1&1&-1&1&-1\\
0&-1&1&-1&1&-1&1\\
-1&0&1&-1&1&1&-1\\
1&0&-1&1&-1&-1&1\\
1&-1&0&1&1&1&1\\
-1&1&0&-1&-1&-1&-1\\
1&-1&-1&0&1&-1&-1\\
-1&1&1&0&-1&1&1\\
-1&-1&1&1&0&-1&-1\\
1&1&-1&-1&0&1&1\\
-1&1&-1&1&1&0&1\\
1&-1&1&-1&-1&0&-1\\
1&1&1&1&1&-1&0\\
-1&-1&-1&-1&-1&1&0\\
0&0&0&0&0&0&0\end{array}$\mbox{ { } }
$\begin{array}{rrrrrrrrrr}
x_1&x_2&x_3&x_4&x_5&x_6&x_7&x_8&x_9&x_{10}\\\hline
0&1&1&1&1&1&1&1&1&1\\
0&-1&-1&-1&-1&-1&-1&-1&-1&-1\\
1&0&-1&-1&-1&-1&1&1&1&1\\
-1&0&1&1&1&1&-1&-1&-1&-1\\
1&-1&0&-1&1&1&-1&-1&1&1\\
-1&1&0&1&-1&-1&1&1&-1&-1\\
1&-1&-1&0&1&1&1&1&-1&-1\\
-1&1&1&0&-1&-1&-1&-1&1&1\\
1&-1&1&1&0&-1&-1&1&-1&1\\
-1&1&-1&-1&0&1&1&-1&1&-1\\
1&-1&1&1&-1&0&1&-1&1&-1\\
-1&1&-1&-1&1&0&-1&1&-1&1\\
1&1&-1&1&-1&1&0&-1&-1&1\\
-1&-1&1&-1&1&-1&0&1&1&-1\\
1&1&-1&1&1&-1&-1&0&1&-1\\
-1&-1&1&-1&-1&1&1&0&-1&1\\
1&1&1&-1&-1&1&-1&1&0&-1\\
-1&-1&-1&1&1&-1&1&-1&0&1\\
1&1&1&-1&1&-1&1&-1&-1&0\\
-1&-1&-1&1&-1&1&-1&1&1&0\\
0&0&0&0&0&0&0&0&0&0\end{array}$\end{small}
\caption{Two screening designs \cite{JN2011}.}\label{tab_bjd}
\end{table}

For $k=7$, the design
is obtained by first folding the points \texttt{0+-+-+-}, \texttt{-0+-++-}, \texttt{+-0++++},
\texttt{+--0+--}, \texttt{--++0--}, \texttt{-+-++0+}, \texttt{+++++-0} and then adding the origin
to total $15$ points.
Under the usual degree reverse lexicographic ordering in \cocoa, we identify the model
with terms:
$1, x_1, x_2, x_3,x_4,
x_5, x_6,$ $x_7, x_6^2, x_7^2, x_2x_7, x_3x_7, x_4x_7, x_5x_7$ and $x_6x_7$. We note
that use of a graded order allows for the inclusion of all terms of degree one before the addition
of terms of second degree, and the total degree of this model (addition of all exponents) is $21$.
If a degree lexicographic order is used, the model remains with the same total degree but
it interchanges one interaction for
a quadratic term:
$1, x_1, x_2, x_3, x_4, x_5,x_6, x_7$,
$x_5^2, x_6^2, x_7^2, x_5x_6, x_4x_7,  x_5x_7, x_6x_7$.

Lexical term orderings work in rather the opposite manner than graded orderings.
For a lexical ordering, then selection of terms
is concentrated in including all terms with $x_7$. As this cannot go further
than $1,x_7,x_7^2$, then term selection allocates all possible terms including $x_6$ until
interaction with $x_6x_7$ appears and term $x_6x_7^2$ can be allocated with $x_7$,
returning to $x_6$ again when required, eventually including terms with $x_5$.
The
model has terms
$1, x_{7}, x_{7}^2, x_{6}, x_{6}x_{7}, x_{6}x_{7}^2,
x_{6}^2, x_{6}^2x_{7}, x_{6}^2x_{7}^2,
 x_{5},
 x_{5}x_{7}, x_{5}x_{6}, x_{5}x_{6}x_{7}, x_{5}^2, x_{5}^2x_{7}$ and its
 total degree is $31$.

Varying term orders over all possible orderings is in general a complex and
expensive task.
In Section \ref{sec_fans}
we discuss and comment on the whole set of models identified by this
seven factor design, when considering all possible term orders.

For $k=10$ the points to be folded are \texttt{0+++++++++}, \texttt{+0----++++},
\texttt{+-0-++--++}, \texttt{+--0++++--},
\texttt{+-++0--+-+}, \texttt{+-++-0+-+-},
\texttt{++-+-+0--+}, \texttt{++-++--0+-},
\texttt{+++--+-+0-} and
\texttt{+++-+-+--0}. The standard term ordering in \cocoa{ } was used to
identify with this design a model of total degree $30$ with constant,
all ten linear terms $x_1,\ldots,x_{10}$, two quadratic terms $x_{9}^2,x_{10}^2$
and eight double interactions
 between $x_{10}$ and one of $x_2,\ldots,x_9$.
%  x_{8}x_{10},  x_{7}x_{10},  x_{6}x_{10},  x_{5}x_{10},  x_{4}x_{10},  x_{3}x_{10},
%  x_{9}x_{10}, x_{2}x_{10}
%
%$1, x_{1}, x_{2}, x_{3}, x_{4}, x_{5}, x_{6}, x_{7}, x_{8}, x_{9}, x_{10},
%x_{10}^2, x_{9}x_{10}, x_{9}^2,  x_{8}x_{10},  x_{7}x_{10},  x_{6}x_{10},  x_{5}x_{10},  x_{4}x_{10},  x_{3}x_{10},  x_{2}x_{10}, $
A lexical term ordering produces a model of much higher total degree ($44$) which contains
monomials in four factors only ($x_6,x_8,x_9$ and $x_{10}$):
$1, x_{10}, x_{10}^2, x_{9}, x_{9}x_{10}, x_{9}x_{10}^2, x_{9}^2,$ $x_{9}^2x_{10}$, $x_{9}^2x_{10}^2, x_{8}, x_{8}x_{10}, x_{8}x_{9}, x_{8}x_{9}x_{10}$, $x_{8}^2, x_{8}^2x_{10}, x_{6}, x_{6}x_{10}$, $x_{6}x_{9}, x_{6}x_{9}x_{10},$ $x_{6}x_{8}$ and $x_{6}x_{8}x_{10}$

}\end{example}

\begin{example}{\normalfont \label{ex_rsm}
Response surface design, non-standard.
Here we take a $16$-point design which is a $5^2$ factorial with all
internal points (a $3^2$ design) removed:
$$\begin{array}{cccc}
  (2,0) & (2,1) & (2,2) & (1,2) \\
  (0,2) & (-1,2) & (-2,2) & (-2,1) \\
  (-2,0) & (-2,-1) &  (-2,-2) & (-1,-2) \\
  (0,-2) & (1,-2) & (2,-2) & (2,-1)
\end{array}$$
The design ideal is generated by the following polynomials
$x_2^5 - 5x_2^3 + 4x_2, x_1^5 - 5x_1^3 + 4x_1,x_1^2x_2^2 - 4x_1^2 - 4x_2^2 + 16$.
It can be shown that for any term ordering, the above polynomials
form a
reduced Gr\"obner basis and thus the design identifies a single
model with terms
$1, x_2, x_2^2, x_2^3, x_2^4, x_1, x_1x_2,$ $x_1x_2^2, x_1x_2^3, x_1x_2^4, x_1^2, x_1^2x_2, x_1^3,$ $x_1^3x_2, x_1^4, x_1^4x_2$.
In Example \ref{ex_ccd} a standard response surface design of the
central composite type is presented.
}\end{example}

\begin{example}{\normalfont \label{ex_rf}\label{ch11ex3}
Regular fraction. Let us take the
resolution III in six variables (all main effects estimated
independently interaction). In classical notation this has defining
contrasts: $\{ABCD,CDEF\}$. Instead of $A,\ldots,F$ we use
indeterminates $x_1,\ldots,x_6$
and selecting one of the four blocks expressed
we have the ideal
$$\langle x_1^2-1,x_2^2-1,x_3^2-1,x_4^2-1,x_5^2-1,x_6^2-1,x_1x_2x_3x_4 -1,
x_3x_4x_5x_6-1\rangle,$$ and setting all polynomials above equal to zero
(simultaneously) gives the design.
The design ideal is created in the following \cocoa{ } code as the sum of the ideal
defining the full factorial design and the ideal defining the desired fraction.
\begin{verbatim}Use T::=Q[x[1..6]];
I:=Ideal([A^2-1|A In Indets()])
   +Ideal(x[1]*x[2]*x[3]*x[4]-1, x[3]*x[4]*x[5]*x[6]-1);
\end{verbatim} The \cocoa{ } command \texttt{QuotientBasis(I);} gives the quotient basis
\begin{verbatim}[1, x[6], x[5], x[5]x[6], x[4], x[4]x[6], x[3], x[3]x[6], x[2], 
 x[2]x[6], x[2]x[4], x[2]x[4]x[6], x[1], x[1]x[6], x[1]x[4],
 x[1]x[4]x[6]]\end{verbatim}
If the confounding relation is desired for a given monomial, this is computed
using the normal form. For example \texttt{NF(x[2]*x[3]*x[6],I);} with output
\texttt{x[1]x[4]x[6]} shows that
over the design, the term $x_2x_3x_6$ is aliased with $x_1x_4x_6$,
equivalently $x_2x_3x_6-x_1x_4x_6\in I(D)$ and thus both terms appear in
the same row of the aliasing Table \ref{tab_alias1}.
The aliasing table is read row-wise e.g. the first row implies that over the
design $1= x_{1}x_{2}x_{3}x_{4}= x_{3}x_{4}x_{5}x_{6}= x_{1}x_{2}x_{5}x_{6}$.
Note that the first column of Table \ref{tab_alias1} corresponds to the quotient basis
computed before, and that the row containing the monomial $1$ has the generators
of the defining contrast.
%\begin{verbatim}
%[1, x[6], x[5], x[4],  x[2], ok x[4]x[5]x[6], in lieu of x_3 x[2]x[5]x[6], in lieu of x_1
%x[2]x[4], x[2]x[4]x[6], x[2]x[6], x[5]x[6],  x[4]x[6], ok
% x[4]x[5], ilo x_3x_6
%x[2]x[5], ilo x_1x_6
%x[2]x[4]x[5], ilo x_1x_4x_6
%x[2]x[4]x[5]x[6]] ilo x_1x_4
%\end{verbatim}

% Table 2
\begin{table}[h]
\begin{center}$ \begin{array}{l|lll}
  1 & x_{1}x_{2}x_{3}x_{4} & x_{3}x_{4}x_{5}x_{6} & x_{1}x_{2}x_{5}x_{6} \\\hline
  x_{1} & x_{2}x_{3}x_{4} & x_{1}x_{3}x_{4}x_{5}x_{6} & x_{2}x_{5}x_{6} \\
  x_{2} & x_{1}x_{3}x_{4} & x_{2}x_{3}x_{4}x_{5}x_{6} & x_{1}x_{5}x_{6} \\
  x_{3} & x_{1}x_{2}x_{4} & x_{4}x_{5}x_{6} & x_{1}x_{2}x_{3}x_{5}x_{6} \\
  x_{4} & x_{1}x_{2}x_{3} & x_{3}x_{5}x_{6} & x_{1}x_{2}x_{4}x_{5}x_{6} \\
  x_{5} & x_{1}x_{2}x_{3}x_{4}x_{5} & x_{3}x_{4}x_{6} & x_{1}x_{2}x_{6} \\
  x_{6} & x_{1}x_{2}x_{3}x_{4}x_{6} & x_{3}x_{4}x_{5} & x_{1}x_{2}x_{5} \\
  x_{1}x_{4} & x_{2}x_{3} & x_{1}x_{3}x_{5}x_{6} & x_{2}x_{4}x_{5}x_{6} \\
  x_{1}x_{6} & x_{2}x_{3}x_{4}x_{6} & x_{1}x_{3}x_{4}x_{5} & x_{2}x_{5} \\
  x_{2}x_{4} & x_{1}x_{3} & x_{2}x_{3}x_{5}x_{6} & x_{1}x_{4}x_{5}x_{6} \\
  x_{2}x_{6} & x_{1}x_{3}x_{4}x_{6} & x_{2}x_{3}x_{4}x_{5} & x_{1}x_{5} \\
  x_{3}x_{6} & x_{1}x_{2}x_{4}x_{6} & x_{4}x_{5} & x_{1}x_{2}x_{3}x_{5} \\
  x_{4}x_{6} & x_{1}x_{2}x_{3}x_{6} & x_{3}x_{5} & x_{1}x_{2}x_{4}x_{5} \\
  x_{5}x_{6} & x_{1}x_{2}x_{3}x_{4}x_{5}x_{6} & x_{3}x_{4} & x_{1}x_{2} \\
  x_{1}x_{4}x_{6} & x_{2}x_{3}x_{6} & x_{1}x_{3}x_{5} & x_{2}x_{4}x_{5} \\
  x_{2}x_{4}x_{6} & x_{1}x_{3}x_{6} & x_{2}x_{3}x_{5} & x_{1}x_{4}x_{5}   \end{array}$
\end{center}\caption{Aliasing table for Example \ref{ch11ex3}.}\label{tab_alias1}
\end{table}

For regular fractions like this case, the effect of different term orderings in the model
means selecting a (possibly) different representative per each row of the aliasing
table. If for instance the ring is defined instead with a lexical ordering with the
command \texttt{Use T::=Q[x[1..6]], Lex;} %was used
%instead of the first line in the above \cocoa{ } code, 
then the model terms enter in a lexical fashion. Ten terms
$1, x_2,   x_4,  x_5, x_6,
x_2x_4,$ $ x_2x_6,  x_4x_6,$ $ x_5x_6, x_2x_4x_6$
of the model coincide with
the model identified above and six terms
$x_1,x_3,x_1x_4,x_1x_6,x_3x_6,x_1x_4x_6$ are replaced by
$x_2x_5x_6,x_4x_5x_6,x_2x_4x_5x_6,x_2x_5,$ $x_4x_5,x_2x_4x_5$. In each
case, the replacement monomial is in the same row.
%x[2]x[5]x[6], in lieu of x_1
%x[4]x[5]x[6], in lieu of x_3
%x[2]x[4]x[5]x[6]] ilo x_1x_4
%x[2]x[5], ilo x_1x_6
% x[4]x[5], ilo x_3x_6
%x[2]x[4]x[5], ilo x_1x_4x_6
}\end{example}

\begin{example}{\normalfont \label{ex_pb}
Plackett-Burman, PB(8).
 Consider the Plackett-Burman design \cite{PB1946} with $8$ points
 in $k=7$ dimensions generated by circular shifts of
the generator \texttt{+++-+--} together with the point
\texttt{+++++++}. %$(1,1,1,1,1,1,1,1)$
With the standard ordering in \cocoa,
we retrieve the usual first order model:
 $1,x_1,x_2,x_3,x_4,x_5,x_6,x_7$. If a lexical term
ordering in which $x_1\succ \cdots\succ x_7$ is used, the model
retrieved is a ``slack" model in only four factors with terms
$1,x_4,x_5,x_6,x_7,x_5x_6,x_5x_7,x_6x_7$.
%[1, x[7], x[6], x[6]x[7], x[5], x[5]x[7], x[5]x[6], x[4]]
}\end{example}

\begin{example}{\normalfont\label{ex_latsq}
Latin Square. It is a straightforward
exercise to code up combinatorial a designs using indicator
variables. Let us take as an example the $4 \times 4$ Graeco-Latin
square derived via the standard Galois field method. The square is
%$$\begin{array}{cccc}
%  A \alpha & B \beta & C \gamma & D \delta \\
%  B \beta & A \alpha & D \delta & C \gamma \\
%  C \gamma & D \delta & A \alpha & B \beta \\
%  D \delta & C \gamma & B \beta  & A \alpha
%\end{array} $$
$$
\begin{array}{cccc}
  A \alpha & B \beta & C \gamma & D \delta \\
  B \gamma & A \delta & D \alpha & C \beta \\
  C \delta & D \gamma & A \beta & B \alpha \\
  D \beta & C \alpha & B \delta  & A \gamma
\end{array}
$$

Coding up the design with $0-1$ indicators: $x_{ij},\; i,j = 1,
\ldots, 4$, where $i$ indexes the factors rows, columns, and Latin,
Greek letters and $j$ the factor ``levels". The design points in
this coding are shown in Table \ref{tab_grlat}.
Using the a graded lexicographic term ordering in \cocoa, the model
identified for the design is
\begin{verbatim}
[1, u[4], u[3], u[2], t[4], t[3], t[2], c[4], c[3], c[2], r[4], 
 r[3], r[2], t[4]u[4], t[4]u[3], t[4]u[2]]
\end{verbatim}
where the factors labelled \texttt{u,t} identify treatments (Latin and
Greek letters) and the factors labelled \texttt{c,r} identify rows and
columns of the design. Note the neat decomposition of model terms that
coincides with the standard analysis of variance for this orthogonal
design:

\begin{center}\begin{tabular}{lr}
Source&d.o.f.\\\hline
Mean&1\\
\texttt{u} (treatment factor 1)&3\\
\texttt{t} (treatment factor 2)&3\\
\texttt{r} (row factor)&3\\
\texttt{c} (column factor)&3\\
interaction \texttt{tu} (error)&3\\\hline
Total&16
\end{tabular}
\end{center}
The interaction between treatment factors (three terms
involving \texttt{tu} above)
is often allocated to the error.
}\end{example}

%Table 3
\begin{table}[h]
$$\begin{array}{c}
(1,0,0,0,1,0,0,0,1,0,0,0,1,0,0,0)\\(1,0,0,0,0,1,0,0,0,1,0,0,0,1,0,0)\\
(1,0,0,0,0,0,1,0,0,0,1,0,0,0,1,0)\\(1,0,0,0,0,0,0,1,0,0,0,1,0,0,0,1)\\
(0,1,0,0,1,0,0,0,0,1,0,0,0,0,1,0)\\(0,1,0,0,0,1,0,0,1,0,0,0,0,0,0,1)\\
(0,1,0,0,0,0,1,0,0,0,0,1,1,0,0,0)\\(0,1,0,0,0,0,0,1,0,0,1,0,0,1,0,0)\\
(0,0,1,0,1,0,0,0,0,0,1,0,0,0,0,1)\\(0,0,1,0,0,1,0,0,0,0,0,1,0,0,1,0)\\
(0,0,1,0,0,0,1,0,1,0,0,0,0,1,0,0)\\(0,0,1,0,0,0,0,1,0,1,0,0,1,0,0,0)\\
(0,0,0,1,1,0,0,0,0,0,0,1,0,1,0,0)\\(0,0,0,1,0,1,0,0,0,0,1,0,1,0,0,0)\\
(0,0,0,1,0,0,1,0,0,1,0,0,0,0,0,1)\\(0,0,0,1,0,0,0,1,1,0,0,0,0,0,1,0)\\
\end{array}$$
\caption{Design points for Graeco-latin design of Example \ref{ex_latsq}.}\label{tab_grlat}
\end{table}

\begin{example}{\normalfont\label{ex_bibd}
Balanced Incomplete Block Design, BIBD.
Consider the balanced incomplete block design with $n=12$ runs
and $t=6$ treatments $t_1,\ldots,t_6$ arranged in
$b=6$ blocks of size two $[t_i,t_j]$ for the following pairs $(i,j)$:
 \[(1,2),(2,3),(3,4),
(4,5),(5,6),(1,6).\] Using the standard term ordering in \cocoa{ }
gives the following model:

\texttt{[1, t[6], t[5], t[4], t[3], t[2], b[6], b[6]t[6], b[5], b[4], b[3], b[2]]}

A similar decomposition to that of Example \ref{ex_latsq} would allocate the interaction
$b_6t_6$ to the residual error with only one degree of freedom. Under a lexical
ordering we retrieve the same model as above. This
result is not extremely surprising given the highly restricted range of
monomial terms for the model for this design. Thus the biggest influence in selection of model
terms is given by the ordering of the indeterminates, also known as initial ordering, see
 \cite{PRW2000}.
}\end{example}

\begin{example}{\normalfont\label{ex_lhs}
Latin Hypercube Sample. Latin hypercubes \cite{MBC1979} are widely used
schemes in the design and analysis of computer experiments. The design
region is often the hypercube $[0,1]^k$ and designs of interest are often
those that efficiently cover the design region.
Latin hypercubes have at least two clear advantages: univariate projections
of the design are uniform and they are simple to generate.

The design $L_1$ with points $(0, 0, 0), (1/5, 1, 4/5), 
(2/5, 3/5, 2/5),$\linebreak $(3/5, 4/5, 1/5),$ $(4/5, 1/5, 1)$ and  $(1, 2/5, 3/5)$ is an example of
randomly generated latin hypercube in $k=3$ dimensions
and  $n=6$ runs. Under the standard term
ordering in \cocoa, the design $L_1$ identifies the model
$1,x_1,x_2,x_3,x_2x_3,x_3^2$.
Experimentally, some latin hypercubes have been found to identify certain types
of models which are of minimal degree called ``corner cut models'', see \cite{OS1999}
also \cite{BMORW2009}.
The design $L_1$ belongs
to such class, and will be discussed further in Section \ref{sec_fans}.

A second example of latin hypercube is $L_2$ with points
$(0, 0, 4/5),$ $(1/5, 1/5, 2/5),$ $(2/5, 2/5, 1)$,
$(3/5, 3/5, 0), (4/5, 4/5, 3/5)$ and $(1, 1, 1/5)$.
Under the same ordering as above, $L_2$ identifies the model
$1,x_2,x_3,x_2x_3,$ $x_3^2,x_3^3$.
%[1, c, c^2, c^3, b, bc]

}\end{example}

\section{Understanding aliasing}\label{ch11_alias}
The algebraic method is not only a way of obtaining candidate models
but it does, we claim, deliver considerable understanding  of the
notion of aliasing. Aliasing is close to the idea of equivalence
used above to define the quotient operation.

Let $I(D)$ be the design ideal and for two polynomials $f,g$ define
$$f(x) \sim_D g(x) $$
to mean
$f(x)=g(x),\; x \in D.$
This is equivalent to $$f(x) - g(x) \in I(D).$$
Again equivalently we have, with respect to a particular monomial ordering
$\prec$,
$$NF(f) = NF(g)$$
We call this {\em algebraic} aliasing.

However, this is not quite the same as the statistical idea of aliasing. It would be enough that $f
= cg$ over the design for some non-zero constant $c$. That is  both $f$
and $g$ should not both be in the same regression model.
We first need a notation to refer to values of a polynomial $f(x)$ on the design expressed
as a vector we write this as
$\mbox{supp}_D(f(x)).$
Then $f(x) \sim_D g(x) $ is equivalent to
$$ \mbox{supp}_D (f(x)) = \mbox{supp}_D(g(x))$$

\begin{definition}
Collections of polynomials $F$ and $G$ are said to be
statistically aliased if
\begin{equation}\mbox{span} \{ \mbox{supp}(f),\; f \in F \} = \mbox{span} \{ \mbox{supp}(g),\; g \in G \}\label{alias2}
\end{equation}
\end{definition}
and let us write this as
$$F \approx_D G$$ %  \label{alias2}
Given that $f(x) = \mbox{NF}(f), x \in D$, we can rewrite \ref{alias2}  as
$$\mbox{span} \{ \mbox{supp}(NF(f)),\; f \in F \} = \mbox{span} \{ \mbox{supp}(NF(g),\; g \in G \}.$$
This means that any aliasing statement is equivalent to one for the normal forms.
For $f \in F$, let $$f = \sum_{\alpha \in L} \theta_{\alpha,f} x^{\alpha},$$
where $L$ is as defined above and depends on the design $D$ and the monomial ordering,
$\prec$. Let $\theta_f$ be the vector of $\theta_{\alpha,f}$ and define  $\theta_{\alpha,g}$, similarly.
Then since the matrix $X$
in non-singular, by construction, we have
$$ F \approx_D G \Leftrightarrow \mbox{span} \{\theta_f, f \in F\} = \mbox{span} \{\theta_g, g \in F\}$$

Thus, statistical aliasing can be thought of in two stages: (i) first reduce to expressing each
polynomial in $F$ and $G$ to its normal form using the algebra then (ii) compare the coefficient subspaces.
In the regular factorial fraction case the normal form of a monomial is itself a monomial, which makes the interpretation
easier, but in the general case it is a polynomial.

We can often we can find the alias classes by inspection, once we
have the normal form. Consider Example 2 and the monomials
$\{x_1^2x_2^2,x_1^4 x_2^4, x_1^6 x_2^6,$ $x_1^8 x_2^8\}$. Then, using
\cocoa{ } the normal forms are, respectively,
$$
\begin{array}{c}
4x^2 + 4y^2 - 16,\;
16x^4 + 16y^4 - 256, \\
320x^4 + 320y^4 - 256x^2 - 256y^2 - 4096, \\
5376x^4 + 5376y^4 - 5120x^2 - 5120y^2 - 65536
\end{array}
$$
We see by inspection that
%\end{array}
$$\{1,x_1^2x_2^2,x_1^4 x_2^4\} \approx_D  \{1, x_1^6x_2^6,x_1^8x_2^8\}.$$
The equivalence continues to all $\{1, x_1^{2k}x_2^{2k},x_1^{2k+1}x_2^{2k+1}\}$.

To retain the link to classical notation we might say that the collection $\{I,A^2B^2,A^4B^4\}$
is aliased with the collection $\{I,A^6B^6,A^8B^8\}$ and we might write
$$\{I,A^2B^2,A^4B^4\} \approx \{I,A^6B^6,A^8B^8\} $$
This arises because $A^2B^2 \approx A^2+B^2-4I$ and  $A^4B^4 \approx A^4+B^4-16$, and both the reduced forms are estimable.

In this example odd terms also pair up. The normal forms of $\{x_1^3 x_2^3,x_1^5 x_2^5, x_1^7,$ $x_2^7,x_1^9 x_2^9\}$ are respectively
$$
\begin{array}{c}
4x^3y + 4xy^3 - 16xy,\; 80x^3y + 80xy^3 - 384xy \\1344x^3y + 1344xy^3 - 6656xy,\;
21760x^3y + 21760xy^3 - 108544xy
\end{array}
$$
So that
$\{1,x_1^3x_2^3,x_1^5 x_2^5\} \approx  \{1, x_1^7 x_2^7,x_1^9x_2^9\},$
and so on, and in classical notation: $\{I,A^3B^3, A^5B^5\} \approx \{I,A^7B^7, A^9B^9\}$.

\section{Indicator functions and orthogonality}

At times it is convenient to see the design $D$ as a  subset of a full factorial design $\mathcal N$.
This is most usual when we start with some basic design, such as a full factorial, and consider a fraction.
We saw such a fraction in the last subsection.
In this case an algebraic description of the fraction is via an indicator function:
$F_D$, rather than a G-basis.
The design ideal of $D$ is unique, what changes are the generating equations we choose to describe it.
These encode different information on $D$.

An indicator function is a {\em single}
additional function which we add to the generators of the ideal of the full factorial design
to form the ideal of $D$. We can write the last example as
$$I(D) = \langle x_1^2-1, x_2^2-1, x_3^2-1, x_1x_2x_3 +1\rangle.$$
The first three terms form the $G$-basis of the full factorial
$\{(\pm 1,\pm 1,\pm 1)\}$.

From the equation $f(x_1,x_2,x_3)=x_1x_2x_3+1=0$ we can deduce the indicator functions of $D$ in $\mathcal N$ as
$g(x_1,x_2,x_3)=\frac{2-f}2=\frac{1}{2}(-x_1x_2x_3+1)$. This takes the value $1$
on the design and $0$ on $\mathcal N\setminus D$. Then, on $D$:
$$ x_1x_2x_3+1=0 \Leftrightarrow g(x_1,x_2,x_3)=1$$

More generally let $\mathcal N$ be the basic, starting  design which is not necessarily a full factorial design,
and let $D\subset \mathcal N$ be a fraction. Fix a monomial order and, via the $I(\mathcal N)$,
construct a vector space basis for interpolation over $\mathcal N$. Then the indicator
function of $D$ interpolates the $0,1$ values as required:
$$
g(x) = \left\{
\begin{array}{l}
  1,\; x \in D \\
  0,\; x \in \mathcal N  \backslash D \\
\end{array}
\right.
$$
In the example above, there is only one basis for interpolation, being $\mathcal N$ a full factorial design: $\{  x_1^{\alpha_1} x_2^{\alpha_2} x_3^{\alpha_3}: \alpha_{i}\in \{0,1\} \textrm{ for } i=1,2,3 \}$
and the indicator function involves only the terms for $\alpha=(0,0,0) $ and $\alpha=(1,1,1)$.

The coefficients of the indicator functions expressed over the interpolation basis embed information on the 'geometric/combinatoric' properties of the fraction.
We exemplify this in the binary case where $\mathcal N$ is the $2^d$  with coding $\{-1,1\}$ \cite{FPR2000}.
For factors with mixed levels a  coding with complex numbers is needed \cite{PR2008}.

Two (square-free) monomials $x^{\alpha}, x^{\beta}$ are said to be are said to be {\em orthogonal} over $D\subset \mathcal N$ if the corresponding columns in the $X$-matrix are orthogonal:
$$\sum_{x\in D'} x^{\alpha}x^{\beta}  = \sum_{x\in D'} x^{\alpha+\beta}= 0.$$
We can express this in terms of the indicator function over $\{-1,1\}^d$ and
write
$$\sum_{x \in \mathcal N} x^{\alpha + \beta}g(x)=0$$
because $g(x)=0$ over $\mathcal N\setminus D$ and  $g(x)=1$  over $D$.
In the example above we want to check that the two-way factors are not orthogonal to the one-way factor. Indeed
$$ \sum_{x \in \mathcal N}  x_1x_2x_3 \frac{2-x_1x_2x_3}2 =
 \sum_{x \in \mathcal N}   \frac{ 2 x_1x_2x_3  -1}2 =
 \sum_{x \in \mathcal N}  x_1x_2x_3 - 4=-4\neq 0
 $$
because $x_1^2=1$ over $\mathcal N$.  It is no coincidence that the coefficient of $x_1x_2x_3$ in the indicator function is not zero. Out of the zero coefficients of $g$ one can deduce the orthogonal (monomial) functions over $D$.

A very practical advantage of the indicator function is that we can take union and intersections of design rather
easily by using Boolean type operations over $D$:
$$g_{D_1 \cap D_2} = g_{D_1}g_{D_2},\; g_{D_1 \cup D_2} =
g_{D_1}+g_{D_2}- g_{D_1}g_{D_2}.$$
Again the zero coefficients of the normal form of $g_{D_1 \cap D_2} $ and $g_{D_1 \cup D_2} $ over the interpolation monomial basis of $\mathcal N$
are informative of the geometry of the intersection and union design.

\section{Fans, state polytopes and linear aberration}\label{sec_fans}

The computations of Gr\"obner basis and model identification with
Gr\"obner basis described in Sections \ref{sec_gb} and \ref{sec_ed}
depend upon the term ordering selected.
Setting a fixed term order
allows the experimenter to put preference over terms which will be
identified by the model, for instance a graded ordering will include
as many terms of order one as possible factors before adding terms
of second degree in the model.
In other instances, the experimenter might be interested in exploring
the range of all models identifiable by the design using algebraic
techniques. For example this would allow assessment of design
properties like estimation capacity \cite{CC2004,CM1998} or the
minimal linear aberration of the design \cite{BMORW2009} and its
general case of non-linear aberration \cite{BLMORWW2008}.
Fan computations have been applied among others, to industrial
experiments \cite{HPRW1999} and systems biology \cite{DJLS2007}.

Some figures of this Section were generated with \texttt{gfan}
 and computations were performed with \cocoa{ } and
\texttt{gfan} \cite{cocoa,gfan}.

%% Figure
\begin{figure}[h]
\begin{center}
\includegraphics[height=10cm]{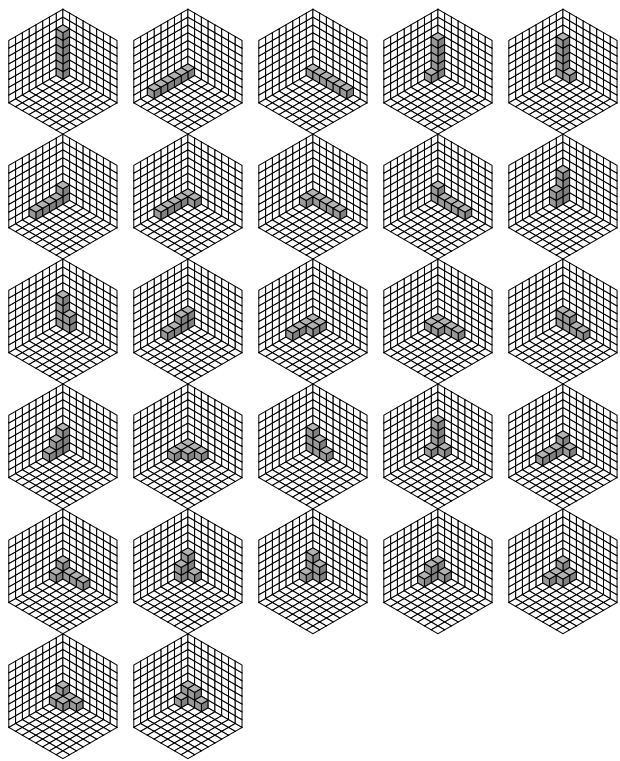}%
\end{center}
\caption{Algebraic fan of design $L_1$ of Example \ref{ex_lhs}.}
\label{fgfan}
\end{figure}

\subsection{The algebraic fan of a design}
Given a design ideal $I(D)$ and ranging over all possible term
orderings, we have a collection of reduced Gr\"obner bases for
$I(D)$. A crucial fact is that despite the infinite number of
different term orderings (excluding the trivial case of one dimension),
this collection of bases has always a finite number of distinct
elements \cite{MR1988}. Associated to this collection of Gr\"obner
bases there is a collection of polyhedral cones, called the Gr\"obner
 fan, and we term the algebraic fan of the design to the collection of
different bases for the quotient ring $R/I(D)$. Note that the algebraic
fan is effectively, a collection of saturated models.

For some relatively simple designs, such as factorial designs, the algebraic fan has
only a single model. The general class of designs with a single model
is called {\em echelon} designs \cite{PRW2000}. However, at
present, computation of the algebraic fan of a design remains an
expensive computation. Reverse
search techniques
are at the core of state-of-the-art software \texttt{gfan} \cite{gfan}.
However, other approaches remain under investigation, such as the
polynomial-time approach based on partial orderings, operations
with matrices and zonotopes \cite{BOT2003,MA2006}.
The well known link between Gr\"obner basis calculations
and linear algebra operations for zero dimensional ideals (i.e. design
ideals) allows these methodologies to be efficient \cite{DHLO2009, L2010}

\begin{figure}[h]
\begin{center}
\includegraphics[height=5cm]{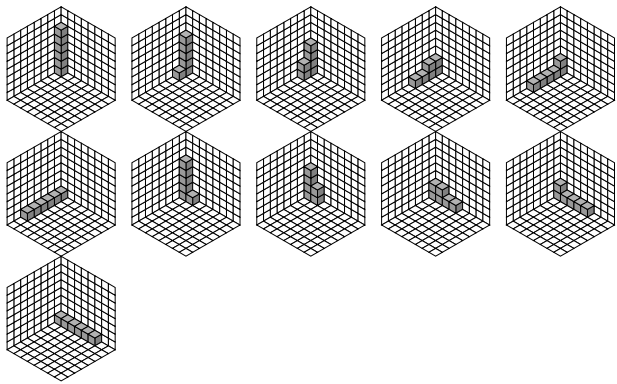}%
\end{center}
\caption{Algebraic fan of design $L_2$ of Example \ref{ex_lhs}.}
\label{fgfan2}
\end{figure}

\begin{example}\label{ex_lhs2}{\normalfont (Continuation of Example \ref{ex_lhs}) The collection
of all models identifiable by the design $L_1$ (algebraic fan of $L_1$) was computed.
Design $L_1$ identifies $27$ different models which can be classified
in only six types of models, up to permutations of variables:
 $1,x_1,x_1^2,x_1^3,x_1^4,x_1^5$ (3 models);
 $1,x_1,x_1^2,x_1^3,x_1^4,x_2$ (6 models);
 $1,x_1,x_1^2,x_1^3,x_2,x_1x_2$ (6 models);
 $1,x_1,x_2,x_1^2,x_1x_2,x_2^2$ (3 models);
 $1,x_1,x_2,x_3,x_1^2,x_1^3$ (3 models) and
 $1,x_1,x_2,x_3,x_1^2,x_1x_2$ (6 models).
 
We say that this fan has a complete combinatorial structure, meaning that
each class of models is closed under permutations of indeterminates,
e.g. if the model $1,x_1,x_1^2,x_1^3,x_1^4,x_1^5$ is in the fan,
so are the models $1,x_2,x_2^2,x_2^3,x_2^4,x_2^5$ and
$1,x_3,x_3^2,$ $x_3^3,x_3^4,x_3^5$, obtained by permuting indeterminates.

The algebraic fan of $L_1$ is depicted in Figure \ref{fgfan}, where
each model is represented as a staircase diagram, with indeterminates
 $x_1,x_2,x_3$ along axes and one small box for each monomial term. The
models are presented by classes following the order described
above (row-wise from top left). For instance, the first diagram shows the
model $1,x_3,x_3^2,x_3^3,x_3^4,x_3^5$, the second is $1,x_1,x_1^2,x_1^3,x_1^4,x_1^5$
and so on.

Now we turn our attention to the other latin hypercube $L_2$. From the design
coordinates we note that this design has complete confounding between $x_1$ and $x_2$
and we should expect a much more limited collection of models. Indeed this design
identifies only $11$ models which are depicted in Figure \ref{fgfan2}.
Only one of the models contains terms with $x_1$ (first from left in
second row); while the rest of the models have
 monomials in $x_2$ and $x_3$. The models can be classified in three classes,
 only one of which is closed under permutation of indeterminates (shown in the
 left column in Figure \ref{fgfan2}).
}\end{example}

%% Figure 2

%% Table 4
\begin{table}[h]
\begin{center}
\begin{tabular}{cclrc}\hline
Simplicial&\multirow{2}{*}{Degree}&\multirow{2}{*}{$HS(s)$}&\multirow{2}{*}{Class size}&Example\\
complex&&&&Vertex/Model\\\hline
\multirow{2}{*}{\psset{unit=8mm}\begin{pspicture}(0,0)(1,1)
\psdots(0.15,0.8)(0.6,0.4)(0.7,0.1)(0.2,0.3)(0.9,0.9)(0.95,0.35)(0.55,0.7)
\end{pspicture}
}&\multirow{2}{*}{$7$}&\multirow{2}{*}{$1+7s$}&\multirow{2}{*}{$1^*$}&$(1,1,1,1,1,1,1)$\\
&&&&$1,x_1,x_2,x_3,x_4,x_5,x_6,x_7$\\\hline
\multirow{2}{*}{\psset{unit=8mm}\begin{pspicture}(0,0)(1,1)
\psdots(0.15,0.8)(0.6,0.7)(0.7,0.1)(0.2,0.3)(0.9,0.9)(0.95,0.35)
\psline(0.9,0.9)(0.95,0.35)
\end{pspicture}
}&\multirow{2}{*}{$8$}&\multirow{2}{*}{$1+6s+s^2$}&\multirow{2}{*}{$105^*$}&$(1,1,1,2,1,2,0)$\\
&&&&$1,x_1,x_2,x_3,x_4,x_5,x_6,x_4x_6$\\\hline
\multirow{2}{*}{\psset{unit=8mm}\begin{pspicture}(0,0)(1,1)
\psdots(0.15,0.8)(0.7,0.1)(0.2,0.3)(0.9,0.9)(0.95,0.35)
\psline(0.95,0.35)(0.9,0.9)(0.15,0.8)
\end{pspicture}
}&\multirow{2}{*}{$9$}&\multirow{2}{*}{$1+5s+2s^2$}&\multirow{2}{*}{$420$\phantom{$^*$}}&$(0,1,2,2,3,0,1)$\\
&&&&$1,x_2,x_4,x_3,x_5,x_7,x_4x_5,x_3x_5$\\\hline
\multirow{2}{*}{\psset{unit=8mm}\begin{pspicture}(0,0)(1,1)
\psdots(0.15,0.8)(0.3,0.25)(0.9,0.9)(0.85,0.35)
\psline(0.85,0.35)(0.9,0.9)(0.15,0.8)(0.85,0.35)
\end{pspicture}
}&\multirow{2}{*}{$10$}&\multirow{2}{*}{$1+4s+3s^2$}&\multirow{2}{*}{$84$\phantom{$^*$}}&$(1,3,0,3,0,3,0)$\\
&&&&$1,x_1,x_2,x_4,x_6,x_2x_4,x_2x_6,x_4x_6$\\\hline
\end{tabular}\end{center}\caption{Summary of the algebraic fan of the
Plackett-Burman design.}\label{tab_pb8}
\end{table}

\begin{example}\label{ex_pb2}{\normalfont (Continuation of Example \ref{ex_pb})
In total there are $610$ different
hierachical models identifiable by the Plackett-Burman design. Those models
belong to $4$ different classes, only two of which are generated by
all permutations of factors. Note that as the design has only two levels in
each factor, the models identified by this design are all multilinear. The
lowest total degree of models is $7$, and the largest total degree is
$10$. See Table \ref{tab_pb8} for details and examples for each class, where
the sign $^*$ refers to a class which is closed under permutation of
indeterminates. The Hilbert Series $HS(s)$ has been included to describe
model terms degree by degree in each class.
}\end{example}

% Table 5
\begin{table}[h]
\begin{center}
\begin{tabular}{cccrc}\hline
\multirow{2}{*}{Class}&Total&\multirow{2}{*}{$HS(s)$}&Class&Example\\
&degree&& size&Vertex/Model\\\hline
\multirow{3}{*}{I}&\multirow{9}{*}{$26$}&\multirow{9}{*}{$\begin{array}{l}1+6s+7s^2\\+2s^3\end{array}$}&\multirow{3}{*}{$24$}&$(5,5,6,2,1,7)$\\
&&&&$1,x_1,x_2,x_3,x_4,x_5,x_6,x_1x_2,x_1x_3,x_1x_6,$\\
&&&&$x_2x_3,x_2x_6,x_3x_6,x_4x_6,x_1x_3x_6,x_2x_3x_6$\\\cline{1-1}\cline{4-5}
\multirow{3}{*}{II}&&&\multirow{3}{*}{$24$}&$(7,1,4,4,3,7)$\\
&&&&$1,x_1,x_2,x_3,x_4,x_5,x_6,x_1x_3,x_1x_4,x_1x_5,$\\
&&&&$x_1x_6,x_3x_6,x_4x_6,x_5x_6,x_1x_3x_6,x_1x_4x_6$\\\cline{1-1}\cline{4-5}
\multirow{3}{*}{III}&&&\multirow{3}{*}{$24$}&$(4,4,6,2,2,8)$\\
&&&&$1,x_1,x_2,x_3,x_4,x_5,x_6,x_1x_3,x_1x_6,x_2x_3,$\\
&&&&$x_2x_6,x_3x_6,x_4x_6,x_5x_6,x_1x_3x_6,x_2x_3x_6$\\\hline
\multirow{3}{*}{IV}&\multirow{6}{*}{$28$}&\multirow{6}{*}{$\begin{array}{l}1+5s+7s^2\\+3s^3\end{array}$}&\multirow{3}{*}{$24$}&$(6,6,2,6,0,8)$\\
&&&&$1,x_1,x_2,x_3,x_4,x_6,x_1x_2,x_1x_4,x_1x_6,x_2x_4,$\\
&&&&$x_2x_6,x_3x_6,x_4x_6,x_1x_2x_6,x_1x_4x_6,x_2x_4x_6$\\\cline{1-1}\cline{4-5}
\multirow{3}{*}{V}&&&\multirow{3}{*}{$24$}&$(4,8,0,8,4,4)$\\
&&&&$1,x_1,x_2,x_4,x_5,x_6,x_1x_2,x_1x_4,x_2x_4,x_2x_5,$\\
&&&&$x_2x_6,x_4x_5,x_4x_6,x_1x_2x_4,x_2x_4x_5,x_2x_4x_6$\\\hline
\multirow{3}{*}{VI}&\multirow{3}{*}{$32$}&\multirow{3}{*}{$\begin{array}{l}1+4s+6s^2\\+4s^3+s^4\end{array}$}&\multirow{3}{*}{$12$}&$(8,0,8,8,0,8)$\\
&&&&$1,x_1,x_3,x_4,x_6,x_1x_3,x_1x_4,x_1x_6,x_3x_4,x_3x_6,$\\
&&&&$x_4x_6,x_1x_3x_4,x_1x_4x_6,x_1x_3x_6,x_3x_4x_6,x_1x_3x_4x_6$\\\hline
\end{tabular}\end{center}\caption{Summary of the algebraic fan of regular
fraction $2^{6-2}$.}\label{tab_f262}
\end{table}

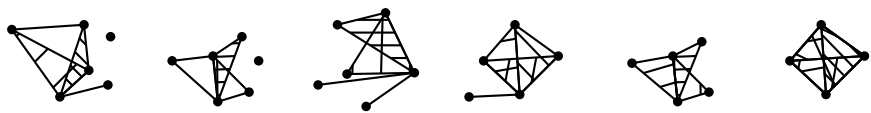
\begin{figure}[h]
\begin{center}
\psset{unit=3.2mm}
\begin{pspicture}(0,0)(6,4)
\psdots(3,0.5)(4.2,1.6)(1,3.3)(4,3.5)(5,1)(5.1,3)
\psline(1,3.3)(4,3.5)\psline(5,1)(3,0.5)
\pspolygon[fillstyle=vlines](3,0.5)(4.2,1.6)(4,3.5)
\pspolygon[fillstyle=hlines,hatchsep=11pt](3,0.5)(4.2,1.6)(1,3.3)
\end{pspicture}
\begin{pspicture}(-2,0)(4,4)
\psdots(1.2,0.3)(1,2.2)(2.5,0.7)(-0.7,2)(2.2,3)(2.9,2)
\pspolygon[fillstyle=vlines,hatchangle=0.4,hatchsep=10pt](1.2,0.3)(1,2.2)(2.5,0.7)
\pspolygon[fillstyle=hlines,hatchangle=1.9](1.2,0.3)(1,2.2)(2.2,3)
\psline(1,2.2)(-0.7,2)(1.2,0.3)
\end{pspicture}
\begin{pspicture}(0,0)(6,4)
\psdots(5,1.5)(1,1)(3,0.1)(3.8,4)(2.2,1.45)(1.8,3.5)
\pspolygon[fillstyle=vlines,hatchangle=-1.4,hatchsep=10pt](3.8,4)(2.2,1.45)(5,1.5)
\pspolygon[fillstyle=hlines,hatchangle=0.4](1.8,3.5)(5,1.5)(3.8,4)
\psline(3,0.1)(5,1.5)(1,1)
\end{pspicture}
\begin{pspicture}(0,0)(6,4)
\psdots(0.9,0.5)(3,0.6)(2.8,3.5)(1.5,2)(4.6,2.2)
\pspolygon[fillstyle=vlines,hatchangle=44.4,hatchsep=7pt](3,0.6)(2.8,3.5)(4.6,2.2)
\pspolygon[fillstyle=vlines,hatchangle=-10.4,hatchsep=9pt](3,0.6)(1.5,2)(4.6,2.2)
\pspolygon[fillstyle=hlines,hatchangle=10.4,hatchsep=10pt](3,0.6)(2.8,3.5)(1.5,2)
\psline(0.9,0.5)(3,0.6)
\end{pspicture}
\begin{pspicture}(-2,0)(4,4)
\psdots(1.2,0.3)(1,2.2)(2.5,0.7)(-0.7,1.9)(2.2,2.8)
\pspolygon[fillstyle=vlines,hatchangle=0.4,hatchsep=9pt](1.2,0.3)(1,2.2)(2.5,0.7)
\pspolygon[fillstyle=hlines,hatchangle=1.9](1.2,0.3)(1,2.2)(2.2,2.8)
\pspolygon[fillstyle=hlines,hatchangle=15.9,hatchsep=6pt](1,2.2)(-0.7,1.9)(1.2,0.3)
\end{pspicture}
\begin{pspicture}(0,0)(6,4)
\psdots(3,0.6)(2.8,3.5)(1.5,2)(4.6,2.2)
\pspolygon[fillstyle=vlines,hatchangle=44.4,hatchsep=4pt](3,0.6)(2.8,3.5)(4.6,2.2)
\pspolygon[fillstyle=vlines,hatchangle=-10.4,hatchsep=13pt](3,0.6)(1.5,2)(4.6,2.2)
\pspolygon[fillstyle=hlines,hatchangle=10.4,hatchsep=10pt](3,0.6)(2.8,3.5)(1.5,2)
\pspolygon[fillstyle=vlines,hatchangle=64.4,hatchsep=12pt](2.8,3.5)(1.5,2)(4.6,2.2)
\pspolygon[fillstyle=vlines,hatchangle=24.4,hatchsep=15pt](3,0.6)(1.5,2)(2.8,3.5)(4.6,2.2)
\end{pspicture}
\end{center}
\caption{Depiction of simplicial models for fan classes I-VI (left to right), design $2^{6-2}$.}\label{fig_262}
\end{figure}

\begin{example}{\normalfont (Continuation of example \ref{ch11ex3})
The fan of the regular $2^{6-2}$ fraction with generators $\{ABCD,CDEF\}$
is of relatively modest size: $132$ models which belong to six equivalence
classes whose size range from $12$ to $24$.
Models range from total degree $26$ to $32$ and none of
the equivalence classes is closed under permutation of indeterminates,
which is not entirely surprising given the regularity of the design.
Despite this apparent fan simplicity, this six classes share only three
different total degrees and Hilbert functions. For instance, three different
model classes share the
same total degree $26$ while other two different model classes have
total degree $28$.
Table \ref{tab_f262} shows a summary of the fan computations for this
design, and Figure \ref{fig_262} shows simplicial representation of
models in each class (vertices refer to single factors, edges to
two factor interactions and so on).
}\end{example}

%% Figure 3

\begin{example}{\normalfont (Continuation of Example \ref{ex_bj})
The algebraic fan of the screening design for seven
factors $k=7$ and $n=15$ runs is a complicated and large object
which nevertheless exhibits in some instances combinatorial symmetry.
The design identifies $18368$ staircase models which can be classified in
$25$ equivalence classes. The class sizes range from $7$ to $2520$, while
total degree of models range from $21$ to $31$.
Six equivalence classes are closed under
permutation of indeterminates, and this includes the classes of models
identified by degree lexicographic ($420$ models) and by degree reverse
lexicographic ($210$ models); examples of models for each ordering were
computed in Example \ref{ex_bj}.
Other $2$ equivalence classes are
almost closed, each can be paired with a other small equivalence class.
}\end{example}

\subsection{State polytope and linear aberration}

The state polytope of $I(D)$ is a geometric object which is associated with
the Gr\"obner fan of $I(D)$ \cite{BM1988,MR1988}. The state polytope is
constructed as the convex hull of state vectors, and each state vector is built
from a model in the algebraic fan by simply adding the exponents of
the model. Aside from a constant, indeed each state vector is the
centroid of the staircase diagram represented by the model and thus
the state polytope is the convex hull of all those centroids.

The state polytope of $I(D)$ encodes information by variables about the
total degree of each model in the fan of design $D$. A simple argument of
linear programming shows that models in the algebraic fan are those that
minimise a simple linear cost function on the weighted degree of the model.
This is the idea of linear aberration defined in \cite{BMORW2009}. This concept
has been generalised to nonlinear cost functions \cite{BLMORWW2008,DHLO2009}.

\begin{example}\label{ex_lhs3}{\normalfont For the latin hypercube design $L_1$, the
state polytope of its design ideal is built with state vectors for
each of the $27$ models ennumerated in Example \ref{ex_lhs2}.
For instance, the model $1,x_1,x_1^2,x_1^3,x_1^4,x_1^5$
has state vector
\[(0,0,0)+(1,0,0)+(2,0,0)+\ldots+(5,0,0)=(15,0,0)\] and as the other
two models in this class are created by permutations of variables,
the same action is performed on the state
vectors so for this class we have three vectors:
$(15,0,0),(0,15,0)$ and $(0,0,15)$.
A similar construction and arguments are used for each model
in the fan of $L_1$ and we have $6$ vectors with permutations
of each of $(10,1,0)$, $(7,2,0)$ and
$(4,2,1)$ ; three permutations for each of $(4,4,0)$ and
$(6,1,1)$.
}\end{example}

There is a special type of polynomial models which are of minimal
weighted degree. These models are termed corner cut staircases \cite{OS1999},
as their exponents can be separated by their complement by a single
hyperplane. The properties of corner cut staircases and their cardinality
have been studied in literature \cite{CRST1999,W2002}.

A design that identifies all corner cut models is termed a generic design,
and automatically a generic design is of minimal linear aberration \cite{BMORW2009}.
The collection of models identified by design $L_1$ (of Examples \ref{ex_lhs}, \ref{ex_lhs2}
and \ref{ex_lhs3}) is the set of all corner cut staircases
for $k=3$, $n=6$ and thus $L_1$ is a generic design.
State polytopes associated with corner cuts and generic designs were described in
\cite{IM2003}.

% Figure 4
\begin{figure}[h]
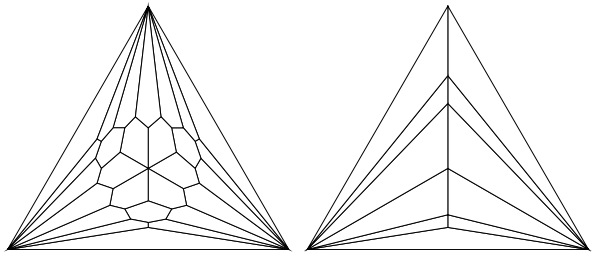
\begin{center}
\includegraphics[scale=0.125]{exg4f.pstex}{ } % to add
\includegraphics[scale=0.125]{exg3f.pstex}
\end{center}\caption{Gr\"obner fan for designs $L_1$ (left) and $L_2$
(right) of Example \ref{ex_lhs}.}\label{fig_lhs}
\end{figure}

In addition to information about degrees of models in the fan, the
state polytope also
encodes information to compute Gr\"obner bases. To each vertex
of the state polytope, a normal cone is associated \cite{Z1995}.
The collection of all those cones is precisely the Gr\"obner fan
of $I(D)$, in the sense that the interior of each full dimensional
cone contains ordering vectors necessary to compute the
Gr\"obner basis (and identify the model) for the corresponding
vertex.

In Figure \ref{fig_lhs}, cones in the fan of state polytopes
for designs $L_1$ and $L_2$ are depicted. As in each case
the tridimensional cones form  a partition of the first orthant,
the figures show a slice of the cones when intersected with the
standard simplex. The diagram for design $L_1$ (left panel) shows
$27$ cells, one for each model. The central symmetry of the diagram
corresponds to symmetry of models under permutation of indeterminates.
Ordering vectors taken from the
same cell will yield the same vertex (and corresponding model).
Now in contrast with generic design $L_1$, design $L_2$
produced the right panel in Figure \ref{fig_lhs}. The diagram
shows still some symmetry, but not central symmetry. This symmetry
reflects the range of models computed for $L_2$ in Example
\ref{ex_lhs2}, where only $11$ models are identifiable by $L_2$,
and ten models are in terms of $x_2$ and $x_3$. The following example
illustrates changes in the fan by addition of one point to the
design.

\begin{figure}[h]
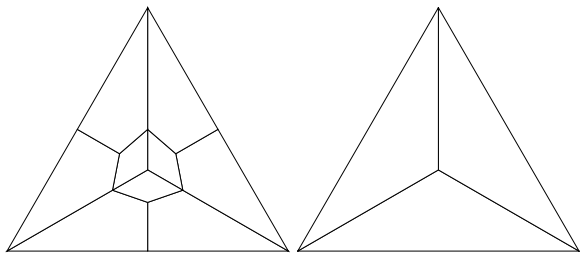

\begin{center}
\includegraphics[scale=0.125]{ccdwof.pstex}{ }%
\includegraphics[scale=0.125]{ccdwif.pstex}
\end{center}
\caption{Algebraic fan of central composite designs in Example \ref{ex_ccd}.}
\label{fgfanccd}
\end{figure}

\begin{example}{\normalfont Response surface design, central 
composite design. \label{ex_ccd}
Consider the central composite design design in three factors
built with axial points at distance $\sqrt 2$ and a full factorial
design with points at levels $\pm 1$. If no point is added to the
origin, this design has $14$ points and a combinatorial algebraic 
fan with $6$
models. The models in the fan belong to only two classes, one with 
monomials
$1, z, z^2, z^3, z^4, y, yz, yz^2, y^2,$ $x, xz, xz^2, xy, xyz$ and the 
other class replaces $z^4$ by $x^2$ above. Addition of
the origin to the previous design has a simplification effect in
the fan, reducing to only $3$ models, while it remains
combinatorial.
The only class of models is created by the list above together
with the monomial $x^2$. See depictions of both fans in Figure 
\ref{fgfanccd}, with the left panel depicting design without origin 
and the right panel after adding the origin.
}\end{example}

\section{Other topics and references}

The algebraic method in the form discussed here can be said to have
started started with \cite{PW1996}, \cite{DS1998}, \cite{D1998} and
the basic ideas were presented in the monograph \cite{PRW2000}. A
short review is \cite{R2009}. More extensive work on the computation
of universal Gr\"obner basis with zonotopes appears in
\cite{BOT2003}.  Applications to designs appear in mixtures
\cite{GWR2001} and  \cite{MNR2007}. Industrial applications were
performed,
perhaps surprisingly early:  \cite{HPRW1999} \cite{PRW2000b}.

For an excellent summary of the wider work in the field of Algebraic
statistics: see \cite{DSS2009}. One important topic omitted from
this chapter, but important for conducting exact conditional test
for contingency tables via Markov Chain Monte Carlo is the
construction of Markov bases; see \cite{HAT2010} \cite{HST2012}
\cite{HT2010} \cite{CR2011}. Important applications to biology,
which continue, are covered in \cite{PS2005}.
Related and of considerable recent
interest is the algebraic study of boundary Exponential models:
\cite{RKA2011} \cite{CP2007} \cite{DS2007}.

Recent work showed the link between minimal aberration models and
the border description of models in terms of Betti numbers of
monomial ideals \cite{MSW2011}, see also extensive references in that
paper.

\end{document}